\documentclass [prd,eqsecnum,nofootinbib]  {revtex4}
\usepackage{amsfonts,amsmath,amscd}
\usepackage{graphicx}
\usepackage{dsfont}

\def\D{\displaystyle{d}}
\def\E{\displaystyle{e}}

\def\SO{\mathsf{SO}}

\begin{document}
\title{%
Kinematics of a relativistic particle with de Sitter momentum space}
\author{Michele Arzano}
\email{marzano@uu.nl} \affiliation{%
Institute for Theoretical Physics and Spinoza Institute,
Utrecht University, Leuvenlaan 4,
Utrecht 3584 TD, The Netherlands}
\author{Jerzy Kowalski-Glikman}
\email{jkowalskiglikman@ift.uni.wroc.pl}\affiliation{Institute for Theoretical Physics,
University of Wroc\l{}aw, Pl.\ Maxa Borna 9, Pl--50-204 Wroc\l{}aw, Poland}
\date{\today}

\begin{abstract}
We discuss kinematical properties of a free relativistic particle with deformed phase space in which momentum space is given by (a submanifold of) de Sitter space.  We provide a detailed derivation of the action, Hamiltonian structure and equations of motion for such free particle. We study the action of deformed relativistic symmetries on the phase space and derive explicit formulas for the action of the deformed Poincar\'e group. Finally we provide a discussion on parametrization of the particle worldlines stressing analogies and differences with ordinary relativistic kinematics.

\end{abstract}
\maketitle

\section{Introduction}
Field theories with curved momentum space have a long history which dates back to the early 60s when they were studied as possible examples of divergence free theories \cite{golfand}.  The idea of curving momentum space had been actually already introduced by Snyder in his classic paper on non-commutative space \cite{Snyder:1946qz}.  In the past decade the topic has received a boost of new interest for different reasons.  On one side certain classes of non-commutative field theories covariant under deformed relativistic symmetries have been shown to posses Fourier transformed counterparts which are functions of group-valued momenta.  On the other hand ``group field theories", in the path integral approach, have been studied  as powerful tools to generate  amplitudes in the spin foam approach to quantum gravity .

Interestingly enough group-valued momenta emerge even in more ``canonical" settings namely when analyzing the behaviour of a relativistic particle coupled to gravity.  Indeed it is well known that the phase space of a standard relativistic particle is geometrically described by the cotangent bundle of Minkowski spacetime. In this case the cotangent bundle is just a product of Minkowski spacetime, the position space, and the dual of its tangent space, the momentum space, which is again isomorphic to $\mathbb{R}^4$. In the case when gravity is present this simple picture becomes more complex. First, gravity usually curves space, resulting in the fact that elements of momentum space become non-commutative, with degree of non-commutativity provided by spacetime curvature.  It turns out however that even in the case when local degrees of freedom of gravity are not present and spacetime is  flat gravity may still influence the form of the phase space of the particle in a nontrivial way. This happens, for example, for gravity in $2+1$ dimensions. In this case it follows from Einstein equations that the Riemann tensor vanishes thus spacetime is flat \cite{Staruszkiewicz:1963zz}, \cite{Deser:1983tn}, \cite{Witten:1988hc} and the only effect of gravity is in the possibility of non-trivial topology of the space-time manifold. If we however couple $2+1$ gravity to point particles the behaviour of the latter becomes highly nontrivial: one can show that taking (topological) degrees of freedom of gravity into account results in effective relativistic particle kinematics with curved momentum space, the picture somehow dual to the particle moving in curved spacetime (see e.g., \cite{Matschull:1997du} and \cite{Schroers:2007ey} for a recent compact review.).

In $3+1$ dimensions the situation is much more complex, and only some partial results have been obtained so far. First of all, contrary to $2+1$ dimensions, four-dimensional gravity is not described by a topological field theory, simply because it clearly possesses local degrees of freedom. However it is possible to formulate gravity as a constrained topological field theory \cite{Smolin:2003qu}, \cite{Freidel:2005ak}, with well defined topological limit, in which the local degrees of freedom are switched off and the spacetime becomes locally the flat Minkowski (in the vanishing cosmological constant limit) with its standard group of spacetime symmetries. It is also possible to couple point particles to gravity formulated in this guise \cite{Freidel:2006hv}, \cite{KowalskiGlikman:2006mu}. One expectes that when the topological limit is taken, the theory is going to exhibit the same kind of phenomena as in $2+1$ dimensional case: the particle action is deformed with the particle's momentum space becoming curved, the curvature provided by the mass scale of the theory, the Planck mass $\kappa$.

In this paper we will investigate properties of  a relativistic particle, whose momentum space is the four-dimensional de Sitter space of curvature $\kappa$, which, in terms of the standard coordinates on five dimensional Minkowski space $P_A$, $A=0,\ldots,4$ is defined as a  four-dimensional surface
\begin{equation}\label{I.1.10}
    -P_0^2 +P_1^2+P_2^2+P_3^2+P_4^2=\kappa^2\, .
\end{equation}
In what follows we will make use of the coordinates $k_\mu$, $\mu=0,\ldots,3$ on this space defined by
 \begin{eqnarray}\label{flatc}
 {P_0}(k_0, \mathbf{k}) &=& \kappa\, \sinh
{{k_0}/\kappa} + \frac{\mathbf{k}^2}{2\kappa}\,
\E^{  {k_0}/\kappa}\, , \nonumber\\
 P_i(k_0, \mathbf{k}) &=&   k_i \, \E^{  {k_0}/\kappa}\, , \label{I.1.9}\\
 {P_4}(k_0, \mathbf{k}) &=& \kappa \cosh
{{k_0}/\kappa} - \frac{\mathbf{k}^2}{2\kappa}\, \E^{ {k_0}/\kappa}\,
.\nonumber
\end{eqnarray}
It should be noticed that the coordinates $k$ are not global and cover only a half of de Sitter space defined by the inequality $P_0(k) + P_4(k) >0$, which turns out to be a group manifold of the group $\sf{AN}(3)$. More details can be found in \cite{Freidel:2007hk}. As explained in \cite{Arzano:2009ci} there are pathologies in the construction of theories if we do not restrict the momentum manifold to $\sf{AN}(3)$ which we are going to do here.

The Casimir for Lorentz transformations acting on the first four coordinates (forming the $\SO(3,1)$ subgroup of the $\SO(4,1)$ group that leaves the quadratic form (\ref{I.1.10}) invariant) is given by
\begin{equation}\label{I.1.21}
{\cal C}=    P_0^2-\mathbf{P}^2-m^2\, ,
\end{equation}
and it follows from (\ref{I.1.9}) that in $k$ coordinates it has the form
\begin{equation}\label{I.1.23}
{\cal C}_\kappa \equiv   4\kappa^2\, \sinh^2\left(\frac{k_0}{2\kappa}\right)-{\mathbf{k}^2}\, e^{k_0/\kappa}=m^2\, .
\end{equation}

These are the prerequisites for our construction of the particle's with de Sitter momentum space action which we carry on in the next Section.  After deriving the action we will briefly describe the construction of the phase space and the Hamtonian analysis of the model.  Then in Section III we turn to the description of the action of deformed symmetries on the phase space and discuss the possible physical interpretation.  Such a discussion, we hope, could be of interest in view of recent debate concerning some alleged paradoxes \cite{Hossenfelder:2010tm}\footnote{For discussion and rebuttal of these claims see the recent analysis reported in \cite{Smolin:2010xa}, \cite{AmelinoCamelia:2010qv}.
}, from which, as it is claimed, the whole class theories with deformations provided by an observer-independent energy scale \cite{Amelino-Camelia:2000ge}, \cite{Amelino-Camelia:2000mn}, \cite{KowalskiGlikman:2001gp}, \cite{Bruno:2001mw}, \cite{Magueijo:2001cr} \cite{Kowalski-Glikman:2006vx} suffers. We hope that the analysis presented in this paper will help to further clarify the issue of internal consistency of this class of theories.

\section{Relativistic particle with curved momentum space}
\label{sec:I.1.3}

Let us turn to the construction of the action principle for a single free particle whose momentum space is half de Sitter space covered by the coordinates (\ref{flatc}) \cite{Girelli:2005dc}.
Noticing that de Sitter space is most simply described by equation (\ref{I.1.10}) it seems convenient to start our investigations in 5 dimensions and then reduce the model one dimension down. Thus the lagrangian of a relativistic particle with de Sitter valued momenta should contain three terms: the standard kinetic term $L= \dot X^A\, P_A$, $A=0,\ldots,4$ with dot denoting derivative over the evolution parameter $\tau$, which makes the unconstrained phase space of the model $5+5$ dimensional, and two constraints: one forcing the momentum space to be de Sitter
\begin{equation}\label{I.1.25}
   {\cal S} \equiv P_AP^A-\kappa^2\approx0\, ,
\end{equation}
 and the second enforcing the mass-shell conditions (\ref{I.1.21}) 
 \begin{equation}\label{I.1.26}
 {\cal C} \equiv P_\mu P^\mu - m^2\approx0\, .
 \end{equation}
Note in passing that there is another natural candidate for the deformed mass-shell condition, namely $P_4-\tilde m \approx 0$, $\tilde m^2 = \kappa^2 + m^2$.  Thus the action we start with has the form
\begin{equation}\label{I.1.27}
    S = \int \D\tau\, P_A\, \dot X^A + \Lambda\, {\cal S} + \lambda\, {\cal C}\, , \quad \dot X^A = \frac{\D X^A}{\D\tau} \, .
\end{equation}
It follows that the Poisson brackets of the basic phase space variables $X^A$ and $P_A$ are canonical, i.e.,
\begin{equation}\label{I.1.28}
    \left\{X^A,P_B\right\}=\delta^A_B
\end{equation}
We will solve explicitly the ${\cal S}$ constraint to obtain the action in terms of the $4+4$ phase space variables with momenta belonging to the $\sf{AN}(3)$ group manifold.  As a first step notice that since ${\cal S}$ is a first class constraint we must gauge fix it, by adding to the model another constraint ${\cal G}$ whose Poisson bracket with ${\cal S}$ does not vanish
\begin{equation}\label{I.1.29}
    \left\{{\cal S},{\cal G}\right\}\neq 0\, .
\end{equation}
After imposing ${\cal G}$ we are dealing with a system with two second class constraints, which therefore describes $4+4$-dimensional phase space. It is convenient to parametrize this reduced phase space with variables $x^\mu$, $k_\mu$ that have vanishing Poisson bracket with the constraints
\begin{eqnarray}
  \left\{x^\mu, {\cal S}\right\} &=& \left\{x^\mu, {\cal G}\right\} \approx0\, , \label{I.1.30}\\
  \left\{k_\mu, {\cal S}\right\} &=& \left\{k_\mu, {\cal G}\right\} \approx0\, ,\label{I.1.31}
\end{eqnarray}
where the last equalities mean that the Poisson brackets are weakly zero, i.e., equal zero on the surface  ${\cal S}={\cal G}=0$. It follows that the Dirac brackets of these variables and arbitrary functions of them is equal to their original Poisson bracket, because they differ just by terms proportional to the brackets (\ref{I.1.30}), (\ref{I.1.31}). One can interpret these equations along with (\ref{I.1.29}) as providing the parametrization of the original phase space in terms of $({\cal S}, {\cal G}, x^\mu, k_\mu)$ with the reduced phase space of interest being defined by condition ${\cal S}={\cal G}=0$.
In order to find the reduced action describing the resulting dynamics of the system with curved momentum space we must choose the gauge fixing function ${\cal G}$, find the expressions for $X^A$, $P_A$ in terms of the variables $x^\mu$, $k_\mu$ satisfying (\ref{I.1.30}), (\ref{I.1.31}) in such a way that both ${\cal S}(X,P)$ and ${\cal G}(X,P)$, when expressed in terms of the new variables, are identically zero, and plug them back to the original action (\ref{I.1.27}).

The first step of the construction is therefore to choose the explicit form of the gauge fixing ${\cal G}$. Here we must recall that we are not interested in the whole de Sitter space as a space of momenta, but only in its $\sf{AN}(3)$ group submanifold defined in addition to the condition ${\cal S}=0$ by the requirement $P_0+P_4>0$.  Thus we have a natural choice for ${\cal G}$ if we impose its Poisson bracket with ${\cal S}$ to be fixed by such condition, namely
\begin{equation}\label{I.1.33}
    \left\{{\cal G},{\cal S}\right\}= P_0+P_4\, ,
\end{equation}
which leads to the following expression
\begin{equation}\label{I.1.32}
    {\cal G}= \frac12\left(X^4-X^0\right) - T
\end{equation}
with $T$ being an arbitrary number.

The momentum sector of the reduced phase space can be straightforwardly described by $k_{\mu}$ given by the coordinates on $\sf{AN}(3)$ given in (\ref{flatc}). Indeed any function of the momenta $P_A$ has vanishing Poisson bracket with ${\cal S}$; as for ${\cal G}$ it has vanishing Poisson bracket with $P_i$, $i=1,2,3$ and with $P_0+P_4=\kappa\, \E^{k_0/\kappa}$ and thus with $k_0$ which can be seen as a function of $P_0+P_4$.\\ 
 
The position part of the reduced phase space can be  constructed as follows. Notice that de Sitter space has a ten parameter group of symmetries, whose action is generated, through Poisson bracket by five dimensional Lorentz transformations, with generators $J_{AB}=\frac1\kappa (X_A\, P_B - X_B\, P_A)$, where we add the prefactor to make $J_{AB}$ dimensionless. One easily checks that indeed
\begin{equation}\label{I.1.34}
    \left\{J_{AB},{\cal S}\right\}= 0\, .
\end{equation}
We can therefore define coordinates using linear combinations of $J_{AB}$ which leave ${\cal G}$ invariant. Since these combinations must give us four coordinates $x^0$, $x^i$ one of whose is a rotation scalar and the remaining three form a three vector we choose
\begin{equation}\label{I.1.35}
    x^0 \equiv J_{04}\, , \quad x^i \equiv J_{0i} + J_{4i} \approx -\frac1\kappa\, (P_0+P_4)\, X^i=-X^i\, \E^{k_0/\kappa}\, ,
\end{equation}
which indeed both have weakly vanishing Poisson bracket with ${\cal G}$. One finds (keeping in mind the changes in sign in raiding and lowering indices)
\begin{equation}\label{I.1.36}
    \left\{x^0,x^i\right\}= \frac1\kappa\, x^i\, , \quad \left\{x^i,x^j\right\}=0\, ,
\end{equation}
i.e. the same algebra structure of the so called $\kappa$-Minkowski space \cite{Majid:1994cy}.\\

To derive the reduced action it remains to plug the expressions for $X^A$ as functions of $x^\mu$ and $k_\mu$ along with (\ref{I.1.9}) into (\ref{I.1.27}). One finds (disregarding the total derivatives and changing the sign)
\begin{equation}\label{I.1.37}
    S^{red} = \int \D\tau\, k_\mu \dot x^\mu - \frac1\kappa\, k_i\, x^i\, \dot k_0 + \lambda \left(P_\mu(k) P^\mu(k) + m^2\right)
\end{equation}
We see that the deformation appears in two places: first we have the term $\frac1\kappa\, k_i\, x^i\, \dot k_0$ which deforms the symplectic structure, and second, we have to do with the deformed on-shell condition, enforced by the Lagrange multiplier $\lambda$, $P_\mu(k) P^\mu(k) + m^2=0$. In the following we will take\footnote{Notice that $\lambda$ is a Lagrange multiplier that is needed first to enforce the on-shell condition, and second to make the action invariant under $\tau$ local reparametrizations (one dimensional general coordinate transformations.) Indeed, if we take $\tau \rightarrow \tau'=f(\tau)$ then assuming that both $x(\tau)$ and $k(\tau)$ are scalars with respect to this transformation, the kinetic term in  $S^{red}$ is invariant. To make the mass-shell term invariant as well we must assume that $\lambda(\tau)$ transforms so that $\D\tau'\, \lambda'(\tau') = \D\tau\, \lambda(\tau)$ i.e., $$
\lambda'(\tau')= \left(\frac{\D f(\tau)}{\D\tau}\right)^{-1}\, \lambda(\tau)\, ,
$$
which means that $\lambda$ is a one dimensional vielbein determinant. This means that we can always use this transformation to make $\lambda$ constant. Although one cannot make it equal 1 in general because the length of the interval
$$
\int_{\tau_1}^{\tau_2} \D\tau\lambda
$$
is invariant, in what follows we will ignore this difficulty and use the gauge $\lambda=1$.} $\lambda =1$ remembering that we must add the on-shell condition to the equations of motion.  Varying $S^{red}$ with respect to $x^\mu$ we get the condition that momenta are conserved
\begin{equation}\label{I.1.38}
    \dot k_\mu = 0\, .
\end{equation}
Varying with respect to momenta we obtain relations between components of four velocity and momenta
\begin{equation}\label{I.1.39}
    \dot x^0 = \frac{1}{\kappa} \, k_i\, \dot x^i + \frac{\partial}{\partial k_0}\left( P_\mu(k) P^\mu(k)\right)\, , \quad \dot x^i = \frac{\partial}{\partial k_i}\, \left(P_\mu(k) P^\mu(k)\right)\, .
\end{equation}
We will return to these equations in the following section.

\subsection{Hamiltonian analysis}

Let us now turn to the Hamiltonian analysis of the action $S^{red}$, (\ref{I.1.37}).  Let us denote by $p_\mu$ the momenta associated with position $x^\mu$ and by $\Pi^\mu$ the ones associated with $k_\mu$. By definition we have
\begin{equation}\label{I.1.40}
    p_\mu \equiv \frac{\partial L}{\partial \dot x^\mu} = k_\mu\, , \quad \Pi^i \equiv \frac{\partial L}{\partial \dot k_i} = 0\, , \quad \Pi^0 \equiv \frac{\partial L}{\partial \dot k_0} = -\frac1\kappa\,  k_i\, x^i\, ,
\end{equation}
and the Poisson brackets between such conjugate variables are by definition
\begin{equation}\label{I.1.41}
    \left\{x^\mu, p_\nu\right\} = \delta^\mu_\nu\, , \quad \left\{k_\mu, \Pi^\nu\right\} = \delta^\nu_\mu\, .
\end{equation}
All the conditions (\ref{I.1.40}) are  constraints, because there are no velocities on the right hand sides. The canonical Hamiltonian reads
\begin{equation}\label{I.1.42}
    H \equiv p_\mu\, \dot x^\mu + \Pi^\mu\, \dot k_\mu - L = -\lambda \left(P_\mu(k) P^\mu(k) + m^2\right)\approx 0
\end{equation}
and is a constraint, as always in the case of the re-parametrization invariant systems.  Eqn.\ (\ref{I.1.40}) defines three sets of constraints
$$
\varphi_\mu \equiv p_\mu-k_\mu\approx0\, , \quad \Psi \equiv \Pi^0+\frac1\kappa\,  k_i\, x^i\approx0\, , \quad \Psi^i \equiv \Pi^i\approx0\, ,
$$
with the following non-vanishing Poisson brackets
$$
\left\{\varphi_0, \Psi\right\} =-1\, , \quad \left\{\varphi_i, \Psi^j\right\}=-\delta_i^j\, , \quad \left\{\varphi_i, \Psi\right\}=-\frac1\kappa\,  k_i\, , \quad \left\{\Psi, \Psi^i\right\}=\frac1\kappa\,  x^i\, .
$$
In order to calculate Dirac brackets we must form a matrix $M$ from the brackets above and then to calculate its inverse matrix, whose components (put in the order $\varphi_0, \varphi_i, \Psi, \Psi^i$) are
$$
M^{-1}=\left(
        \begin{array}{cccc}
          0 & \frac1\kappa\, x^i\,\, & 1\,\, & -\frac1\kappa\, k_i \\
          -\frac1\kappa\, x^i\,\, & {0} & 0 & \delta_i^j \\
          -1 & 0 & 0 & 0 \\
          \frac1\kappa\, k_i & - \delta_i^j \,\, & 0 & {0} \\
        \end{array}
      \right)
\, .
$$
Then the non-vanishing Dirac bracket on reduced phase space with coordinates $(x^\mu, k_\nu)$ are
\begin{equation}\label{I.1.43}
   \left\{ x^0, x^i\right\}_D = - \left\{ x^0, \varphi_0\right\}\left(M^{-1}_{\varphi_0,\varphi_j}\right)\left\{ \varphi_j, x^i\right\}= \frac1\kappa\, x^i\, ,
\end{equation}
\begin{equation}\label{I.1.44}
   \left\{ x^0, k_0\right\}_D = - \left\{ x^0, \varphi_0\right\}\left(M^{-1}_{\varphi_0,\Psi}\right)\left\{ \Psi, k_0\right\}=1\, ,
\end{equation}
\begin{equation}\label{I.1.45} \left\{ x^i, k_j\right\}_D = - \left\{ x^i, \varphi_k\right\}\left(M^{-1}_{\varphi_k,\Psi^l}\right)\left\{ \Psi^l, k_j\right\}=\delta^i_j\, ,
\end{equation}
\begin{equation}\label{I.1.46} \left\{ x^0, k_j\right\}_D = - \left\{ x^0, \varphi_0\right\}\left(M^{-1}_{\varphi_0,\Psi^l}\right)\left\{ \Psi^l, k_j\right\}=-\frac1\kappa\, k_j\, ,
\end{equation}
with the other brackets being trivial, which is the desired classical counterpart of the $\kappa$-deformed phase space \cite{Lukierski:1993wx}.  At this point it will be useful to pause for a moment to discuss the structures we have encountered so far from an algebraic viewpoint.  By construction the ``momentum" variables $\{k_{\mu}\}$ are coordinate functions on the group $AN(3)$.  They are also generators of translations and as such can be viewed as a (trivial) Lie algebra with Lie bracket given by their Poisson brackets.  Being coordinate functions on the group $AN(3)$ they have a natural non-abelian composition rule which turns their space into a group.  The space-time coordinates we chose above, due to their non-trivial Poisson brackets,  can be be seen as coordinate functions on such group.  At the same time due to the ``pairing" $\left\{ x^{\mu}, k_{\nu}\right\}_D =\delta^{\mu}_{\nu}$ the co-ordinates $\{x^{\mu}\}$ span the (non-trivial) Lie algebra {\it dual} the one of translation generators.  In both cases the non-trivial Poisson and Lie brackets are {\it characteristic signature} left by the curvature of momentum space on the ``dual" space spanned by the ${x^{\mu}}$.

\section{Symmetries and Noether charges}

The action $S^{red}$, (\ref{I.1.37}) is, by construction, invariant under action of deformed, infinitesimal, global spacetime symmetries: translations and Lorentz transformations, which together form a ten-parameters algebra. Here we will find an explicit form of such transformations and the associated Noether charges.\\
Consider first time translation generated by $k_0$. Since $k_0$ has a nontrivial bracket with $x^0$ only we find that
\begin{equation}
   \delta x^0 = a^0\, , \quad \delta x^i = 0\, , \quad \delta k_\mu =0\,.
\end{equation}
For an arbitrary spacial translation generated by $a^i\, k_i$ we have instead
\begin{equation}
    \delta x^0 =  \frac{1}{\kappa} k_i\, a^i\, , \quad \delta x^i = a^i\, , \quad \delta k_\mu =0
\end{equation}
so  the action will be invariant under the following transformations 
\begin{equation}\label{I.1.47}
    \delta x^0 = a^0 + \frac{1}{\kappa} k_i\, a^i\, , \quad \delta x^i = a^i\, , \quad \delta k_\mu =0 \, .
\end{equation}
To find Noether conserved charges associated with these transformations we use the usual trick: we first make the parameters local, i.e., $\tau$ dependent and then read off the Noether charges $Q^{trans}$ from the general expression\footnote{The proof goes as follows. Let the infinitesimal transformation with parameter $a$ leave the action $S$ invariant, $\delta S =0$. Let us denote by $\phi(\tau)$ collectively all variables of the model. Then
$$\delta S =\int_{\tau_1}^{\tau_2} \D\tau\left(\frac{\partial L}{\partial\phi}\, \delta \phi + \frac{\partial L}{\partial\dot\phi}\, \frac{\D}{\D\tau}\,\delta \phi\right)=\int_{\tau_1}^{\tau_2} \D\tau\frac{\D}{\D\tau}\left( \frac{\partial L}{\partial\dot\phi}\, \delta \phi\right)\, ,
$$
where in the last equality we use equations of motion. Since $\tau_1$ and $\tau_2$ are arbitrary, and $\delta \phi = a f(\phi)$ with some $f$ and $\tau$-independent $a$, the Noether charge $Q=\frac{\partial L}{\partial\dot\phi}\,  f( \phi)$ is time independent, i.e., conserved. But from inspection of the middle term in the formula above it follows that if we allow $a$ to be $\tau$-dependent the only term containing $\dot a$ is $\dot a\, Q$.}  $\delta L = \dot a^\mu\, Q_\mu$. One checks that
\begin{equation}\label{I.1.48}
    Q^{trans}_\mu = k_\mu\, ,
\end{equation}
which are obviously time independent by virtue of equations of motion,  (\ref{I.1.38}).\\

Since the action $S^{red}$ is manifestly invariant under space rotations it is straightforward to write down the infinitesimal rotations with parameter $\theta_i$
\begin{equation}\label{I.1.49}
    \delta x^i = \theta_j\, \epsilon^{ji}{}_l\, x^l\, , \quad \delta k_i = \theta_j\, \epsilon^{j}{}_i{}^l\, k_l\, , \quad \delta x^0=\delta k_0=0\, .
\end{equation}
The associated Noether charges are just the components of angular momentum
\begin{equation}\label{I.1.50}
    M^i=\epsilon^{ij}{}_l\, x^l\, k_j
\end{equation}
and again by using equations of motion (\ref{I.1.38}), (\ref{I.1.39}) one easily checks that they are conserved.\\

Let us finally turn to the invariance of the action $S^{red}$ under infinitesimal boosts parametrized by $\xi_i$. To do that we must first find the transformation rules for components of the momenta $k_\mu$. To this end it is sufficient to notice that it follows from the form of the original action (\ref{I.1.27}) that the spacetime components of the momenta $P$ transform under boosts as four-vectors, to wit
\begin{equation}
    \delta_\xi P_0 = \xi^i\, P_i\, , \quad \delta_\xi P_i = \xi_i\, P_0\, .
\end{equation}
Using this and the explicit expressions for $P$ as functions of $k$ (\ref{I.1.9}) by straightforward calculation we find
\begin{equation}\label{I.1.51}
   \delta_\xi k_i = \xi_i \left[\frac\kappa2\left(1-\E^{-2k_0/\kappa}\right)+\frac{\mathbf{k}^2}{2\kappa}\right]- \frac{k_i\, \xi\cdot k}{\kappa} \, , \quad \delta_\xi k_0 = \xi \cdot k \, ,
\end{equation}
with $\xi \cdot k =\xi^i\, k_i$. Imposing invariance of the action $S^{red}$ one finds that positions must transform like
\begin{equation}\label{I.1.52}
    \delta_\xi x^i = - \xi^i\, x^0 + \frac1\kappa\, x^i \, \xi \cdot k - \frac1\kappa\, k^i \, \xi \cdot x
\end{equation}
and
\begin{equation}\label{I.1.53}
     \delta_\xi x^0 = -\xi \cdot x+ \frac1\kappa\, x^0 \, \xi \cdot k +\xi \cdot x\left[\frac12\left(1-\E^{-2k_0/\kappa}\right) + \frac{\vec{k}^2}{2\kappa^2}\right]\, .
\end{equation}
The conserved charges associated with boosts are
\begin{equation}\label{I.1.54}
    N_i=-k_i\, x^0 - x_i \left[\frac\kappa2\left(1-\E^{-2k_0/\kappa}\right) + \frac{\vec{k}^2}{2\kappa}\right]\, .
\end{equation}
Notice that the transformation laws for positions (\ref{I.1.52}) and (\ref{I.1.53}) can be rewritten in the following simple form
\begin{equation}\label{I.1.55}
    \delta_\xi x^i = - \xi^i\, x^0 - \frac1\kappa\, \epsilon^{ijk} \xi_j \, M_k
\end{equation}
and
\begin{equation}\label{I.1.56}
     \delta_\xi x^0 = -\xi \cdot x + \frac1\kappa\, \xi \cdot N\, .
\end{equation}
It can be checked by explicit calculations that the charge (\ref{I.1.54}) generates Lorentz transformations through Poisson bracket.

\section{Finite deformed Poincar\'e transformations and $\kappa$-Poincar\'e group}

In this section we address the problem of integration of infinitesimal Lorentz transformations derived in the previous section. In this way we will find $\kappa$-Poincar\'e group action\footnote{Quite unfortunately the name ``$\kappa$-Poincare group'' is being used sometime to denote the Hopf algebra dual to the $\kappa$-Poincare algebra \cite{Kosinski:1994ng}.}, which describes the finite transformations corresponding to Poisson bracket representation of $\kappa$-Poincar\'e algebra.

Let us start with Lorentz transformations of momenta. For simplicity we choose our coordinates in such a way that the boosts acts along the first axis, $\xi^i=(\xi, 0,0)$. Instead of calculating the finite boosts $k_\mu(\xi)$ directly it is more convenient to start with the $P_A$ variables that transform as  components of four-vector and a scalar. Thus we have
\begin{equation}\label{1}
    P_0(\xi) = P_0^{(0)}\cosh\xi + P_1^{(0)}\sinh\xi \, , \quad P_1(\xi) = P_1^{(0)}\cosh\xi + P_0^{(0)}\sinh\xi \, ,
\end{equation}
 \begin{equation}\label{2}
     P_2(\xi) = P_2^{(0)}\, , \quad P_3(\xi) = P_3^{(0)}\, , \quad P_4(\xi) = P_4^{(0)} = \kappa^2+m^2\, ,
\end{equation}
where $P_A^{(0)}$ denote the initial values of the components.

Using these equations and relations (\ref{I.1.9}) one can calculate the expressions for $k_\mu(\xi)$, which have been first obtained in \cite{Bruno:2001mw}. In what follows we will not need these explicit formulas.

Our main goal is to derive the action of finite boosts and translations on position variables. Let us start with boosts. The transformation laws (\ref{I.1.55}) and (\ref{I.1.56}) can be rewritten as differential equations for $\xi$ dependence
\begin{eqnarray}\label{2.1}
                    \frac{\D}{\D\xi}\, x^0(\xi) &=&  - x^1(\xi) +\frac{1}{\kappa} N_1(\xi)\nonumber \\
                      \frac{\D}{\D\xi}\, x^1(\xi) &=& - x^0(\xi) \nonumber  \\
                      \frac{\D}{\D\xi}\, x^{2} (\xi) &=& \frac1\kappa\, M_3(\xi)\, , \quad  \frac{\D}{\D\xi}\, x^{3} (\xi) = - \frac1\kappa\, M_2(\xi)\\
\end{eqnarray}
To solve these equations we must first find out what is the $\xi$ dependence of the relevant components of rotation and boost charges. Since the generator of boost is $N_1$ from $\D/\D\xi N_1 \equiv\{N_1, N_1\}=0$ we see that $N_1=n_1$ is just a constant. Similarly, the first component of the rotation charge $M_1(\xi)=m_1$ is constant. As for the other components we have the equations                             
\begin{eqnarray}                                                       
\frac{\D}{\D\xi}\, M_2(\xi) &=& - N_3(\xi) \, , \quad \frac{\D}{\D\xi}\,M_3(\xi) = N_2(\xi) \, ,\nonumber \\
      \frac{\D}{\D\xi}\, N_2(\xi) &=& -M_3(\xi)\, , \quad \frac{\D}{\D\xi}\,N_3(\xi) = M_2(\xi) \, .\nonumber
 \end{eqnarray}
Denoting the initial values with lowercase letters we have
\begin{equation}\label{3}
    \begin{split} M_2(\xi) &= m_2 \cosh \xi + n_3\sinh \xi
\\ M_3(\xi) &= m_3 \cosh\xi - n_2\sinh \xi \\ N_2(\xi) &= n_2 \cosh \xi - m_3\sinh\xi \\ N_3(\xi) &= n_3\cosh\xi + m_2\sinh\xi.
\end{split}
\end{equation}
 Substituting these expressions to (\ref{2.1}) we find
 \begin{equation}\label{4}
    \begin{split} x_0(\xi) &= x_0(0) \cosh \xi + x_1(0) \sinh \xi
- \frac{n_1}\kappa\, \sinh\xi,
\\ x_1(\xi) &= x_0(0) \sinh \xi + x_1(0) \cosh \xi - \frac{n_1}\kappa\, \cosh \xi + \frac{n_1}\kappa\,
\\ x_2(\xi) &= x_2(0) -\frac{n_2}\kappa\, + \frac{n_2}\kappa\, \cosh\xi - \frac{m_3}\kappa\,\sinh\xi
\\ x_3(\xi) &= x_3(0) - \frac{n_3}\kappa\, + \frac{n_3}\kappa\, \cosh\xi + \frac{m_2}\kappa\,\sinh\xi. \end{split}
 \end{equation}
Using (3) these equations can be rewritten in a simpler form, similar to the standard Lorentz transformations, to wit
 \begin{equation}\label{4a}
    \begin{split} x_0(\xi) &= x_0(0) \cosh \xi + \left(x_1
- \frac{n_1}\kappa\right)(0)\, \sinh\xi,
\\ \left(x_1 - \frac{n_1}\kappa\right)(\xi) &= x_0(0) \sinh \xi + \left(x_1 - \frac{n_1}\kappa\right)(0)\, \cosh \xi
\\ \left(x_2 - \frac{n_2}\kappa\right)(\xi) &= \left(x_2 -\frac{n_2}\kappa\right)(0)
\\ \left(x_3 - \frac{n_3}\kappa\right)(\xi) &= \left(x_3 -\frac{n_3}\kappa\right)(0) \, . \end{split}
 \end{equation}
It follows that for a free particle with given values of boost charge $n_i$ the set of four numbers $(x^0, {\cal X}^i)$, ${\cal X}^i = x^i - n^i/\kappa$ behaves as a set of components of a standard Lorentz four-vector.  Notice also that with the help of (\ref{I.1.50}) and (\ref{I.1.54}) we can express the $n_i$ and $m_i$ constants in terms of the initial positions and components of (conserved) momentum $k$. It should be stressed that, contrary to the action of the standard Lorentz group, there is a nontrivial action in transverse direction, which is suppressed by the deformation scale $\kappa$.

\section{Discussion}

Let us now turn to the discussion of some consequences of the deformed Poincar\'e transformations derived in the preceding section. The problem we would like to address is the following question: suppose we have two free particles whose worldlines cross for one observer, say Alice. Do they cross for another Lorentz boosted observer, Bob? This problem have been discussed first (in a slightly different guise) in \cite{Schutzhold:2003yp}, and the argument of this paper has been rebutted in \cite{Arzano:2003da}. Recently discussion on this issue has been sparked by the paper \cite{Hossenfelder:2010tm} and the extensive discussion of this issue has appeared in the follow-up papers \cite{AmelinoCamelia:2010qv}, \cite{Smolin:2010xa}, \cite{Smolin:2010mx}.

Suppose we use the $\kappa$-Minkowski coordinates $x$ introduced above to parametrize events in spacetime, or in the geometrical language, to provide coordinates. Then, as the calculation presented in the Appendix shows, if we take two worldlines whose crossing point coordinates are $(T,X_i)$ for Alice, for Lorentz boosted Bob, in general, the coordinates of {\em all} points on the worldlines {\em do not} coincide. In other words it would seem that
 \begin{center}
 {\em while Alice sees the worldline cross, Bob does not.}
 \end{center}

Let us argue now that this conclusion (taken as a strong point in \cite{Hossenfelder:2010tm} and implicitly accepted in \cite{AmelinoCamelia:2010qv}, \cite{Smolin:2010xa}, \cite{Smolin:2010mx}) does not seem to be the correct description of the state of affairs.

The first thing that we would like to note is that in {\it ordinary} classical mechanics there are two equivalent ways to describe the phase space of a free relativistic particle.  The first more common approach is simply to take as the configuration space the ``range" space of the coordinates of a particle (Minkowski space) and define the (unreduced) phase space as the cotangent bundle of such configuration space.  The physical (on-shell) 
phase space will be given by a six-dimensional sub-manifold of the unreduced phase space whose coordinates parametrize geodesics in Minkowski space.  For the alternative description one starts from an unreduced phase space given by {\it the dual of the Poincar\'e algebra}.  For a spinless free particle the physical phase space is then given by the six-dimensional submanifold defined by the orbits of the Poincar\'e group\footnote{Such approach offers also the most general formulation of a relativistic particle's phase space since encompasses the case of spinning particle and naturally leads to a description of the quantized system and its coupling to gauge fields.} on such algebra (for details see e.g. \cite{Carinena:1989uw} and references therein) defined by hypersurfaces with vanishing Pauli-Lubanski vector and given mass.  In this case the (physical) momenta are naturally associated to the (linear) momenta and the physical positions are defined using the energy of the particle and the boost charges $x_{i}=n_i/p_0$ (for a pedagogical discussion see \cite{Bacry:1988gz}).

In a deformed symmetry setting it is by no mean obvious (as the ongoing debate in the literature shows \cite{Hossenfelder:2010tm, AmelinoCamelia:2010qv, Smolin:2010xa, Smolin:2010mx}) what natural definition of the configuration space of a free relativistic particle one should adopt.  Indeed there is no a priori reason why one should pick the particular set of $\kappa$-Minkowski coordinates as ``physical" coordinates of a particle since (as discussed above) they just happen to describe a dual space to the translation sector of deformed algebra of symmetry but have no other compelling role that makes them desirable candidates to describe a particle's wordline.   Indeed since the equivalent of  ``picture one" above for the phase space  is now missing it seems sensible to resort to the second option and define a particle's worldline using a combination of Poincar\'e momenta.  Looking at the rules governing transformations of positions $x$ under boosts (\ref{I.1.55}), (\ref{I.1.56}) one easily sees that the source of the problems is the fact that positions belonging to different worldlines transform in different way because the outcome of transformation depends on the momentum that the particle carries.  Intuitively we want to construct new positions, linear combinations of $x$, such that their Lorentz transformations do not depend on momentum. Since the boost generators have the standard Poisson bracket it is convenient to parametrize momentum with the help of $P_\mu$ coordinates (\ref{I.1.9}), which transform as components of a standard Lorentz vector.  This is precisely what one would expect following the more general description of phase space in terms of Poincar\'e momenta as described above.

More concretely we characterize every worldline by six parametrs: the components of boost $n_i$ and a tangent vector whose components can be identified with those of the momentum $P_\mu$ (notice that since the particles are on shell the $P_0$ component can be computed from $P_0^2 - P_i^2=m^2$. Thus the set of worldlines (physical phase space) is, as expected, six dimensional. We can now use coordinates in such space to describe coordinates of events in spacetime. Let us first define three dimensional space by mean of the phase space coordinates $\zeta_i = n_i/P_0$ . Intuitively these coordinates span the space slice at time $\zeta_0=0$. Now take a subset of worldlines which are characterized by $P_i=0$, $P_0=M$, which are just worldlines corresponding to particles at rest with respect to the origin. Now take any worldline with $n_i=0$, i.e. that passes through the ``origin" at the initial time. We can define a time coordinate $\zeta_0$ as follows. Let at some event our worldline with $n_i=0$ cross the worldline of a static particle characterized by a given $n_i^{stat} = M \zeta_i$. Then the time is given be the relation $P_0 \zeta_i + P_i \zeta_0=0$, where $\zeta_i$ is known from the boost components of the static particle. Clearly this procedure provides spacetime coordinates analogous to the standard ones, with the usual special relativistic transformation rules.  Thus at the classical one particle level the kinematics of particle with group valued momenta can be described with usual special relativistic tools.\\

One may wonder where the deformation is gone if the particle kinematics is just that of special relativity? To answer this question we must recall the discussion at the end of Section II.  Even though classically a description in terms of undeformed covariant coordinates is available the curvature of momentum space is still ``hiding" in the dual space of translation generators under the form of non-trivial brackets.  Such non-trivial bracket is ultimately responsible \footnote{A non-trivial bracket on the dual algebra of translation generators corresponds to a non-vanishing ``co-commutator" on the algebra of translations which is nothing but the classical limit of the co-product which describes the (non-symmetric) action of symmetry generators on quantum multiparticle states.} for a series of non-symmetric behaviours of the space-time symmetry generators which affect quite crucially the definition of particle, antiparticle and multi-particle states of the corresponding quantum theory \cite{Arzano:2009ci, Arzano:2009wp}.  Thus the physical effects of deformation seem to show up fully only in the quantum theory and it is a challenge for future studies\footnote{Along this line of thought another possibility to solve the non-locality puzzle can be contemplated (such option was also independently considered by L.\ Smolin (private communication)). To see it one has to realize that in the construction above there is one important hidden assumption. Namely we implicitly assumed that free particles worldlines are independent and that the Noether charges corresponding to the multi-particles system are just a sum of the single-particle ones. Such assumption is very natural in the case of special relativity.  In the deformed case however the presence of the deformation parameter makes it possible to construct topological interactions between particles, similar to those present, and well understood, in the case of 2+1 gravity (see e.g., \cite{Bais:1998yn}, \cite{Bais:2002ye}, \cite{Schroers:2007ey} and references therein). In this case the Noether charges of multi-particles systems are not just sums of the single-particles one, but instead they correspond to the co-product structure of Hopf algebra of spacetime symmetries. We will not dwell on this possibility here leaving the detailed discussion to future work.} to establish what, if any, kind of physical effects of the deformation could be revealed at the classical level.
 
\begin{acknowledgments}
We would like to thank Lee Smolin for many discussions and encouragement in carrying out this project.   JKG is supported in part by research projects  NN202318534 and Polish Ministry of Science and Higher Education grant 182/N-QGG/2008/0. He thanks
 Prof. Renate Loll for her hospitality during his visit to University of Utrecht, where this work started.  MA is supported by a Marie Curie Intra-European Fellowship. MA would also like to thank the Perimeter Institute for Theoretical Physics for hospitality while this work was being completed.  Research at Perimeter Institute for Theoretical Physics is supported in part by the Government of Canada through NSERC and by the Province of Ontario through MRI.
\end{acknowledgments} 
 
\appendix*
\section{Non-locality in $x$ coordinates}

Following \cite{AmelinoCamelia:2010qv} we find it convenient to describe the worldline of a particle as a set of points in spacetime for which the boost charge $N_i$ is constant. Take $X^i$ and $T$ the coordinates of the point where the two wordlines cross (for Alice) and denote $t = x^0$. Then the spacetime points belonging to the particle worldline satisfy
\begin{equation}\label{8}
    k_i\, (t(\tau) -T) + E\, (x_i(\tau) -X_i) = 0 \, ,
\end{equation}
where
\begin{equation}\label{9}
    E \equiv \left[\frac\kappa2\left(1-\E^{-2k_0/\kappa}\right) + \frac{\mathbf{k}^2}{2\kappa}\right]\, .
\end{equation}

Since $E = \E^{-k_0/\kappa}\, P_0$ (cf.\ (\ref{I.1.9})), we see that for massless particles for which $P_0 = |\mathbf{P}|$ the velocity is
\begin{equation}\label{10}
   v= \left|\frac{x_i(\tau) -X_i}{t(\tau) -T}\right| = \frac{|\mathbf{k}|}{E}=\frac{|\mathbf{P}|}{P_0}=1\, ,
\end{equation}
(see \cite{Daszkiewicz:2003yr} and \cite{Girelli:2005dc}.). Notice in passing that in terms of $k$ and $E$ the mass shell relation reads
\begin{equation}\label{11}
    E^2 - \mathbf{k}^2 = m^2\, \E^{-2k_0/\kappa}\, .
\end{equation}

Let us now turn to the non-locality issue raised in \cite{Hossenfelder:2010tm} and further discussed in \cite{Smolin:2010xa} and \cite{AmelinoCamelia:2010qv}. Take two worldlines of massless particles that cross, for an observer Alice, at the spacetime event with coordinates $X^i$ and $T$. Thus the worldlines are described by the equations
\begin{equation}\label{12}
 \begin{split} k_i^{(1)}\, (t^{(1)}(\tau) -T) + E^{(1)}\, (x_i^{(1)}(\tau) -X_i) &= 0 \, ,\\ k_i^{(2)}\, (t^{(2)}(\tau) -T) + E^{(2)}\, (x_i^{(2)}(\tau) -X_i) &= 0 \, . \end{split}
\end{equation}
where $E^{(1/2)} = |\mathbf{k}^{(1/2)}|$ because both particles are massless by assumption, and it is clear that they cross at $T$, $X$. These worldlines are seen by Bob as
\begin{equation}\label{13}
 \begin{split} k'{}_i^{(1)}\, (t'{}^{(1)}(\tau) -T'{}^{(1)}) + E'{}^{(1)}\, (x'{}_i^{(1)}(\tau) -X'{}_i^{(1)}) &= 0 \, ,\\ k'{}_i^{(2)}\, (t'{}^{(2)}(\tau) -T'{}^{(2)}) + E'{}^{(2)}\, (x'{}_i^{(2)}(\tau) -X'{}_i^{(2)}) &= 0 \, , \end{split}
\end{equation}
and in general they {\em do not} cross, because the point $T'{}^{(1)}\neq T'{}^{(2)}$,  $X'{}_i^{(1)}\neq X'{}_i^{(2)}$   as a result of the fact that the points belonging to the two worldlines do not transform in the same way. To see this explicitly, let in the Alice's frame the coordinates of the crossing point be $(T, \vec X) = (0, X,0,0)$ and let the particle $(1)$ moves along the second axis $\mathbf{k}^{(1)} = (0,k^{(1)}, 0)$, while the particle $(2)$ along the third $\mathbf{k}^{(2)} = (0,0,k^{(2)})$.  The components of the boost and angular momentum vectors for the particles are (since these are the conserved quantities i.e., independent of the worldline point, it is sufficient to calculate them at the crossing point)\begin{eqnarray}
 \mathbf{n}^{(1)} = (|\mathbf{k}^{(1)}|\, X,k^{(1)}\, T, 0) &\,\,\,& \mathbf{n}^{(2)} = (|\mathbf{k}^{(2)}|\, X,0,k^{(2)}\, T) \, , \\
\mathbf{m}^{(1)} = (0,0, k^{(1)}\, X)     &\,\,\,&  \mathbf{m}^{(2)} = (0,- k^{(2)}\, X, 0)\, .
\end{eqnarray}
Writing the worldline equation as seen by Alice in the form
\begin{equation}\label{14}
    E^{(I)} x^{(I)}_i + k^{(I)}_i t^{(I)} = n^{(I)}_i
\end{equation}
We find that in Bob's boosted frame they have the form
\begin{equation}\label{15}
    E'{}^{(I)} x'{}^{(I)}_i + k'{}^{(I)}_i t'{}^{(I)} = n'{}^{(I)}_i
\end{equation}
with (see (\ref{3}) 
\begin{equation}\label{16}
   \mathbf{n}'{}^{(1)} = (|\mathbf{k}^{(1)}|\, X, k^{(1)}(T\cosh\xi - X\sinh\xi),0)\, , \quad \mathbf{n}'{}^{(2)} = (|\mathbf{k}^{(2)}|\, X,0, k^{(2)}(T\cosh\xi - X\sinh\xi))\, .
\end{equation}
We still need explicit expressions for transformed energy and momentum, which can be found straightforwardly using the fact that the $P_0$, $P_i$ transform as components of a Lorentz four-vector (\ref{1}), (\ref{2}), to wit
\begin{equation}\label{17}
 \begin{split}
 E'= \kappa\, \frac{E\cosh\xi +k_1\sinh\xi}{E\cosh\xi +k_1\sinh\xi+ e^{-k_0/\kappa}\sqrt{\kappa^2 +m^2}}\, ,\\
k_1'= \kappa\, \frac{k_1\cosh\xi +E\sinh\xi}{E\cosh\xi +k_1\sinh\xi+ e^{-k_0/\kappa}\sqrt{\kappa^2 +m^2}}\, ,\\
 k_{2/3}'= \kappa\, \frac{k_{2/3}}{E\cosh\xi +k_1\sinh\xi+ e^{-k_0/\kappa}\sqrt{\kappa^2 +m^2}}\, ,\end{split}
\end{equation}
where on-shell, for massless particles
\begin{equation}\label{18}
    e^{-k_0/\kappa}=1-\frac{| \mathbf{k}|}\kappa
\end{equation}
Substituting these into (\ref{15}) and denoting 
$$
D = \frac{| \mathbf{k}|}\kappa\, (\cosh\xi -1) +1
$$
 we find that the worldlines in Bob's frame take the form
 \begin{eqnarray}
t'{}^{(1)} \sinh\xi + x'{}^{(1)} \cosh\xi = XD^{(1)} &\qquad& t'{}^{(2)} \sinh\xi + x'{}^{(2)} \cosh\xi = XD^{(2)} \nonumber \\
t'{}^{(1)}  + y'{}^{(1)} \cosh\xi = (T\cosh\xi - X\sinh\xi) D^{(1)}&& y'{}^{(2)}=0 \label{19}\\
z'{}^{(1)}=0 && t'{}^{(2)}  + z'{}^{(1)} \cosh\xi = (T\cosh\xi - X\sinh\xi) D^{(2)}\, . \nonumber
\end{eqnarray}
It is clear that if $D^{(1)}\neq D^{(2)}$ i.e., if $k^{(1)}\neq k^{(2)}$ the worldlines do not cross anymore.

For completeness consider now the case, in which the second particle moves along the first axis, towards the origin. We have
\begin{equation}\label{20}
    \vec n^{(2)} = (| \mathbf{k}^{(2)}|( X- T),0,0) \, ,\quad \vec m^{(2)}=(0,0,0)
\end{equation}
and
\begin{equation}\label{21}
    \vec n'{}^{(2)} = (|\mathbf{k}^{(2)}|( X- T),0,0)
\end{equation}
so that
\begin{equation}\label{22}
    x'{}^{(2)}- t'{}^{(2)} = (X-T)\left( e^{-\xi} + \frac{|\mathbf{k}^{(2)}|}\kappa(1-e^{-\xi})\right)\, ,\quad  y'{}^{(2)}=z'{}^{(2)}=0\, .
\end{equation}
Again, for Bob, the worldlines are missing each other.  It can be also shown directly that analogous result holds if Bob is a translated observer.

\end{document}